\documentclass[aps,prl,amsmath,amssymb,twocolumn,preprintnumbers,nofootinbib]{revtex4}
\pdfoutput=1
\usepackage[utf8]{inputenc} %utf8
\usepackage{graphicx}
\graphicspath{{./figures/}}
\usepackage{url}
\usepackage[bookmarks, pagebackref=false]{hyperref}
\usepackage[usenames,dvipsnames]{xcolor}
\definecolor{orange}{cmyk}{0,0.5,1,0}
\definecolor{rossoCP3}{cmyk}{0,.88,.77,.40}
\definecolor{graa}{rgb}{0.8,0.8,0.8}
\definecolor{blaa}{rgb}{0.2,0.2,0.6}
		\hypersetup{
			colorlinks, 
			bookmarksopen, 
			bookmarksnumbered,
			citecolor=blaa, 		%color of links to bibliography
			linkcolor=rossoCP3,	%color of internal links
			urlcolor=rossoCP3,			%color of external links 
			}
\usepackage{amsthm}
\usepackage{bm}% bold math
\usepackage{bbm}
\usepackage{pxfonts}

\usepackage{amsmath,amssymb,amsfonts}
\usepackage{color}
\usepackage{float}
\usepackage{hyperref}
\usepackage[Symbolsmallscale]{upgreek}
\usepackage{amsmath}
\usepackage{amsfonts}
\usepackage{amssymb,dsfont}
\usepackage{graphicx}
\usepackage{amssymb}
\usepackage[vcentermath]{youngtab}
\usepackage[all]{xy}
\usepackage{pstricks}
\usepackage{dsfont}%
\usepackage{bbold}

\usepackage{placeins}
\usepackage{xspace}
\usepackage{cancel} % to make the barred text notation

\usepackage{slashed}
\usepackage[caption=false, labelformat=simple, listofformat=subsimple, labelfont=default, margin=5pt, justification=raggedright]{subfig}

	\makeatletter
		\renewcommand{\p@subfigure}{}
	\makeatother

\usepackage{natbib}

\usepackage{feynmf}

\usepackage{braket}

\newcommand{\beq}{\begin{eqnarray}}
\newcommand{\eeq}{\end{eqnarray}}

\newcommand{\bmp}{\noindent\begin{minipage}{16cm}}
\newcommand{\emp}{\end{minipage}\vskip 7mm} % 7mm untightened

\newcommand   \cO {\mathcal{O}}

\def\lsim{\mathrel{\rlap{\lower4pt\hbox{\hskip1pt$\sim$}}
    \raise1pt\hbox{$<$}}}                % less than or approx. symbol
\def\gsim{\mathrel{\rlap{\lower4pt\hbox{\hskip1pt$\sim$}}
    \raise1pt\hbox{$>$}}}                % greater than or approx. symbol

\begin{document}
%%%%%%%%%%%%%%%%%%%%%%%%%%%%%%%%%%%%%%%%%%%%%%%%%%%%%%%%%%%%%%%%%%%%%%%%%%%

\title{Charging the $O(N)$ model}
\author{Oleg {\sc Antipin}$^{\color{rossoCP3}{\clubsuit}}$}
\email{oantipin@irb.hr} 
\author{Jahmall {\sc Bersini}
$^{\color{rossoCP3}{\clubsuit}}$}
\email{jbersini@irb.hr} 
\author{Francesco {\sc Sannino} $^{\color{rossoCP3}{\diamondsuit},\color{rossoCP3}{\heartsuit}}$}
\email{sannino@cp3.dias.sdu.dk}
\author{Zhi-Wei Wang $^{\color{rossoCP3}{\diamondsuit}}$}
\email{wang@sdu.dk}
\author{Chen Zhang $^{\color{rossoCP3}{\spadesuit}}$}
\email{czhang@cts.nthu.edu.tw}
\affiliation{{ $^{\color{rossoCP3}{\clubsuit}}$ Rudjer Boskovic Institute, Division of Theoretical Physics, Bijeni\v cka 54, 10000 Zagreb, Croatia}\\{ $^{\color{rossoCP3}{\diamondsuit}}$\color{rossoCP3} {CP}$^{ \bf 3}${-Origins}} \& the Danish Institute for Advanced Study {\color{rossoCP3}\rm{Danish IAS}},  University of Southern Denmark, Campusvej 55, DK-5230 Odense M, Denmark. \\\mbox{ $^{\color{rossoCP3}{\heartsuit}}$Dipartimento di Fisica “E. Pancini”, Università di Napoli Federico II | INFN sezione di Napoli}\\ \mbox{Complesso Universitario di Monte S. Angelo Edificio 6, via Cintia, 80126 Napoli, Italy.}\\
{$^{\color{rossoCP3}{\spadesuit}}$ Physics Division, National Center for Theoretical Sciences, Hsinchu, Taiwan 300}}

\begin{abstract}
 We determine, for the first time,  the scaling dimensions of a family of fixed-charge operators stemming from  the critical $O(N)$ model in $4-\epsilon$ dimensions to the leading  and next to leading order terms in the  charge expansion but to all-orders in the coupling.  We test our results to the maximum known order in perturbation theory while determining higher order terms.  
\\
[.3cm]
{\footnotesize  \it Preprint: RBI-ThPhys-2020-09, CP$^3$-Origins-2020-04 DNRF90,  NCTS-TH/2005}

\end{abstract}

\maketitle
 
\small
%\section{Introduction}  
The discovery of the Higgs heralds a new era in our understanding of fundamental interactions. It crowns the Standard Model of particle physics as one of the most successful theories of nature while simultaneously opening new avenues tailored at gaining a deeper understanding of the ultimate laws of nature. 

One of the most striking features of the Standard Model is its near scale invariant nature. In fact, at the classical level, the only two operators that explicitly violate scale invariance are the Higgs mass and the cosmological constant. It is therefore natural to start investigating the dynamics of theories of fundamental interactions around their scale invariant limit. Scale invariance is highly intertwined with conformal symmetry which leads to powerful constraints on the dynamics of the theory at hand. It is therefore useful to organise quantum field theories around their conformal limit. Therefore in our analysis we shall use it as a tool to access important information about the theory. 

In Reference \cite{Badel:2019oxl} the authors employed a semiclassical approach to determine the scaling dimensions of the fixed charge $\phi^n$ operator, $\Delta_{\phi^n}$, in the $U(1)$ scalar $\phi^4$-model at the Wilson-Fisher (WF) fixed point.  The Standard Model Higgs is, however, described by a non-abelian $O(4)$ model, up to gauge and Yukawa interactions. This calls for generalising the approach to non-abelian theories which, as we shall see, is quite involved.  

We will, therefore, consider $O(N)$ theories and determine the scaling dimensions of a family of fixed-charge operators to the leading  and next to leading order terms in the  charge expansion but to all-orders in the coupling. 
Our work builds upon the pioneering idea of using the large-charge limit~\cite{Hellerman:2015nra,Alvarez-Gaume:2019biu} to gain relevant information about conformal dynamics~\cite{Orlando:2019skh}.   We test our results to the maximum known order in perturbation theory while determining higher order terms.  We plan to generalise our results to generic gauge-Yukawa theories that are the backbones of any known theory of fundamental interactions.

\section{ $O(N)$ at fixed charge}\label{O(N)}

As mentioned in the introduction, in Ref.~\cite{Badel:2019oxl} (see also \cite{Arias-Tamargo:2019xld}) the $U(1)$ model was investigated using a semiclassical method in order to compute the scaling dimensions of the fixed-charge $\phi^n$ composite operator, $\Delta_{\phi^n}$, in the  $\frac{\lambda}{4} \phi^4$-model, with $\lambda$ the self-coupling.  This was performed by analysing the conformal theory  at the Wilson-Fisher (WF) fixed point obtained by going away from four dimensions via $d = 4-\epsilon$ with $\epsilon$ positive and tiny. Using the operator-state correspondence \cite{Cardy:1984rp,Cardy:1985lth} one can use the conformal map of the theory at the WF fixed point  on a cylindrical gravitational background to determine $\Delta_{\phi^n}$ via the  expectation value of the evolution operator $e^{-HT}$ on an arbitrary state $\ket{\psi _n}$ carrying charge $n$ with $H$ the Hamiltonian of the system and $T$ the time coordinate. 

To make this explicit, consider the path integral formula for $\bra{\psi_n}e^{-HT}\ket{\psi_n }$
\begin{align}\label{path}
\bra{\psi_n}e^{-HT}\ket{\psi_n } = \frac{1}{\mathcal{Z}} \int  D \chi_i  D \chi_f 
\psi_n(\chi_i)\psi_n^*(\chi_f) \int^{\rho=f, \chi=\chi_f}_{\rho=f, \chi=\chi_i}  D \rho  D \chi e^{-S} \nonumber\\
\end{align}
%\begin{eqnarray}\label{path}
%\bra{\psi_n}e^{-HT}\ket{\psi_n }= D \chi_i  D \chi_f \
%\psi_n(\chi_i)\psi_n^*(\chi_f) \nonumber  
%\int^{\rho=f,~ \chi=\chi_f}_{\rho=f, ~\chi=\chi_i}  D \rho  D \chi e^{-S}\,, \nonumber \\
%\end{eqnarray}
where $\mathcal{Z}$ normalizes the vacuum-to-vacuum transition amplitude while the wave-functional
\begin{equation}
\psi_n(\chi)=\exp\left( \frac{i\,n}{R^{d-1}\Omega_{d-1}}\int d\Omega_{d-1}\chi\right)
\end{equation}
fixes the charge of the initial and final states to $n$. Here $\rho$ and $\chi$ are modulus and phase of the complex scalar field, while $f$ is a constant value. For small values of the quartic coupling, $\lambda$, this path-integral can be computed via the saddle-point method. The remarkable upshot of  \cite{Badel:2019oxl} is that, similarly to the large-$N$ ’t Hooft expansion in gauge theories, the result can be organized as a double expansion in $\lambda$ and $\lambda n$ with $\lambda n$ fixed.  In other words, the anomalous dimension of $\phi^n$ can be written as
\begin{equation}\label{eq:semiclassics}
\Delta_{\phi^n}=\sum_{\kappa=-1}^{\infty}\lambda^{\kappa}\Delta_\kappa(\lambda n) \ .
\end{equation}
where each of the functions $\Delta_\kappa(\lambda n)$ are computed semiclassically to all-orders in the fixed 't Hooft $\lambda n$ coupling. A similar analysis, entirely in four dimensions because of the existence of an ultraviolet interacting perturbative fixed point, appeared earlier in \cite{Orlando:2019hte}  for non-abelian gauge-Yukawa theories but concentrating on the large charge limit.  

\subsection*{$O(N)$ model setup and ground state}
Here, inspired  by the fact that the Standard Model Higgs is described by an $O(4)$ non-abelian model,  we move to analyse the non-abelian massless $O(N)$ vector model described below

\begin{equation} \label{action}
    \mathcal{S}=\int d^d x \left(\frac{(\partial \phi_i)^2}{2}+\frac{(4\pi)^2 g_0}{4!}(\phi_i\phi_i)^2\right) \ .
\end{equation}

In $d= 4-\epsilon$ this theory features an infrared WF fixed point for the renormalized coupling $g$ and its value at the $3$-loop level in the MS scheme reads \cite{H. Kleinert and V. Schulte-Frohlinde}
\begin{widetext}
\begin{eqnarray} \label{WFFP}
g^*(\epsilon)= \frac{3 \epsilon}{8+N}+\frac{9 (3N+14) \epsilon^2}{(8+N)^3}+\frac{ \epsilon^3}{(8+N)^5} \left[\tfrac{3}{8}(4544+1760 N+110 N^2-33 N^3)- 36 \ \zeta(3)(N+8)(5N+22) \right]+\mathcal{O}(\epsilon^4) \ .
\end{eqnarray}
\end{widetext}
 In the $O(N)$ vector model with even (odd) $N$ we can fix up to $n=\frac{N}{2}$ $\left(\frac{N-1}{2}\right)$ charges, which is the rank of the $O(N)$ group. We fix $k\leq n$ of these charges and write the path integral expression to determine the ground state energy of this charge configuration on a cylinder. % \emph{A priori}, one expects all the charges $\bar Q_i$ to enter in such expression but we will show that actually this is not the case.
  From now on, we focus on the even-$N$ case, since the odd case is similar, and rewrite the action in terms of $n$ complex field variables
\begin{align}\label{complex}
  \varphi_1 &= \frac{1}{\sqrt 2} \left(\phi_1 + i \phi_ 2 \right) = \frac{1}{\sqrt 2} \sigma_1 \ e^{i \chi_1}\,, & \\ \varphi_2 &= \frac{1}{\sqrt 2} \left( \phi_3 + i \phi_4\right) = \frac{1}{\sqrt 2} \sigma_2 \ e^{i \chi_2}\,, & \\ \varphi_3 &=\dots \ \ .
\end{align}
At the WF fixed point $g^*$ we map the action onto the cylinder, $\mathbb{R}^d \to \mathbb{R} \times S^{d-1}$, which now reads
\begin{equation}
        \mathcal{S}_{cyl} = \int d^d x \sqrt{g} \left(g_{\mu\nu}\partial^{\mu} \bar \varphi_i  \partial^{\nu} \varphi_i + m^2 \bar \varphi_i \varphi_i + \frac{(4 \pi)^2 g_0}{6} (\bar \varphi_i \varphi_i)^2\right) \ .
\end{equation}
A mass term appears $m^2 = \left(\tfrac{d-2}{2 R}\right)^2$, stemming from  $R$ the radius of the sphere. 
Notice that the procedure above is merely employed to ease the computation.   

The charges are fixed using $k$ constraints $Q_i = \bar Q_i$, where $\{\bar Q_i\}$ is a set of fixed constants. Clearly, $\varphi_i$ ($\bar \varphi_i$) has charge $\bar Q_i = 1 \ (-1)$. The solution of the EOM with minimal energy is spatially homogeneous and it is given by
\begin{equation} \label{SolEOM}
  \begin{cases}
\sigma_i =  A_i\, \ , \ \ \chi_i =- i \mu t \, & i=1,\dots,k \,,\\
\varphi_{k+j} = 0, & j=1,\dots,n-k \, .
\end{cases}
\end{equation}
As  pointed out in \cite{Alvarez-Gaume:2016vff}, a striking consequence of such homogeneous solution is the presence of a single chemical potential $\mu$, even if the charges $\bar Q_i$ are all different.  The parameters $A_i$ and $\mu$ are fixed by the EOM and by the expression for the Noether charge as
\begin{equation} \label{total}
    \mu^2-m^2 = \frac{(4 \pi)^2}{6}g_0 v^2  \qquad \quad \frac{\bar Q}{\text{vol.}}=\mu v^2 \qquad \quad \text{vol.}= 2 \pi^2 R^3 \ ,
\end{equation}
where we have defined
\begin{equation}
  v^2 \equiv \sum_{i=1}^k A_i^2  \quad \qquad  \bar Q \equiv \sum_{i=1}^k \bar Q_i 
\end{equation}
with $\bar Q$ the sum of the charges.

%The presence of a single chemical potential has consequences for our path integral expression for $\bra{\psi_n}e^{-HT}\ket{\psi_n}$. This can be understood by analyzing the symmetry breaking pattern induced by the charge fixing. 
The presence of a single chemical potential leaves the $O(2n-2k)\times U(k)$ symmetry of the original $O(2n)$ symmetry unbroken \cite{Alvarez-Gaume:2016vff}.
Then, the vacuum of the theory spontaneously breaks $U(k)$ to $U(k-1)$. In fact it is possible to rotate the ground state to 
\begin{equation}\label{rot}
\frac{1}{\sqrt 2} (A_1,..., A_k, 0,...,0) \longrightarrow \big(  \underbrace{0, ...,0}_{k-1}\,,\, \tfrac{v}{\sqrt 2} \,,\, \underbrace{0,...,0}_{n-k} \big)
\,. 
\end{equation} 
Stemming from the considerations above the saddle point computation is organized as  single coupling 't Hooft expansion in $g^* \bar Q$ similar to the abelian case. The sum of the charges act as a single charge, which is a welcome simplification leading to 
\begin{equation} \label{finale}
    \bra{\psi_{\bar Q}}e^{-HT}\ket{\psi_{\bar Q}} =   \frac{1}{\mathcal{Z}} \int^{\scriptstyle \sigma_{N/2}=v }_{\scriptstyle \sigma_{N/2} = v} D^n \sigma \ D^n \chi \ e^{- \mathcal{S}_{eff}}
    \end{equation}
where 
\begin{eqnarray}
 \label{O2n_Lagrangian}
\mathcal{S}_{eff} =&& \int^{T/2}_{-T/2} d t \ \int d\Omega_{d-1} \left(\frac{1}{2}\partial \sigma_i \partial \sigma_i + \frac{1}{2} \sigma_i^2 (\partial \chi_i \partial \chi_i)\right. \nonumber \\  && \left.+ \frac{m^2}{2} \sigma_i^2 +\frac{(4 \pi)^2}{24}g_0 (\sigma_i \sigma_i)^2 + \frac{i} {\text{vol.}} \  \bar Q  \ \dot \chi_{N/2} \right) \ .
    \end{eqnarray}
The sums over $i$ go from $1$ to $n = N/2$, i.e. we have fixed the maximum number of charges $k=n$. %Notice that one could also consider the independent charges $i\bar Q_i  \dot \chi_{i}/\text{vol.}$, since 
%\begin{eqnarray}
%\mathcal{S}_{eff} =&& \int d\tau \ d\Omega_{d-1} \left(\frac{1}{2}\partial \sigma_i \partial \sigma_i + \frac{1}{2} \sigma_i^2 (\partial \chi_i \partial \chi_i)\nonumber \\ \right.  && \left.+ \frac{m}{2} \sigma_i^2 +\frac{(4 \pi)^2}{24}g_0 (\sigma_i \sigma_i)^2 + \frac{i} {\text{vol.}} \  \bar Q_i  \ \dot \chi_{i} \right) 
%\end{eqnarray}
%being linear in the field it does not change the LHS of \eqref{finale}.
In conclusion, the scaling dimension at the fixed point of the smallest dimension operator carrying a total charge $\bar Q$ assumes the form
\begin{equation} \label{TQ}
\Delta_{O_{\bar Q}} = E_{O_{\bar Q}}R = \sum_{j=-1}^{\infty} g^{*j}\Delta_j(g^* \bar Q) \ .
\end{equation}
Here $E_{O_{\bar Q}}$ is the ground state energy and $R$ the radius of the sphere. 

\subsection*{Fixed charge operators}

In CFT with internal global symmetries, operators organize themselves into multiplets transforming according to irreducible representations of the symmetry group. Within any one such multiplet, component operators are further distinguished by their charge configurations, namely the value of their charges associated with the Cartan generators. Component operators of different charge configurations do not mix under renormalization. Nevertheless, by virtue of the Wigner-Eckart theorem, they necessarily have the same scaling dimension.

We would like to compute scaling dimensions of the lowest-lying operators corresponding to some fixed charge configuration via a semi-classical expansion, and compare the results to ordinary perturbation theory whenever possible.In a massless theory, operators with different engineering dimensions do not mix. We therefore consider the minimal classical scaling dimension (MCSD) for a given charge configuration. Let's start by assuming  the generic charge configuration to be $[m]=(m_1,m_2,...,m_{N/2})$, with the $m_i$’s representing the charge associated with the $i$th Cartan generator. Without loss of generality, we suppose all $m_i$’s to be positive. Then the fixed-charge operator with the MCSD is $\mathcal{O}_{[m]}\equiv\prod_{i=1}^{i=N/2}(\varphi_i)^{m_i}$ with $\varphi$ complex. If some $m_i$ were to be negative, they would correspond to replace $\varphi_i$ with $\bar{\varphi_i}$. $\mathcal{O}_{[m]}$ must be a tensor operator living in the traceless fully symmetric subspace of $O(N)$ transformations, which corresponds to an irreducible representation and therefore has a definite scaling dimension. $\mathcal{O}_{[m]}$ is fully symmetric simply because it is a product of commuting  scalar field. Furthermore, it is traceless and it turns out to be the MCSD operator with this charge configuration. 
  
We, therefore, arrive at the conclusions that operators that have the same total charge and MCSD all belong to the same irreducible representation of  $O(N)$ and thus have the same scaling dimension.  

We therefore identify such operator to be the $\bar{Q}$-index traceless symmetric tensor  $T_{\bar{Q}} \equiv T_{i_1...i_{\bar{Q}}}^{({\bar{Q}})}$. The latter can be represented as ${\bar{Q}}$-boxes  Young tableaux with one row. 
%Pictorially, the fully symmetric space is
%\begin{eqnarray}
%\label{symmON}
 % \underbrace{ {\small \Yvcentermath1  \yng(3)}...\small \Yvcentermath1  \yng(1)}_{m}+ \underbrace{ {\small \Yvcentermath1  \yng(2)}...{\small \Yvcentermath1  \yng(1)}}_{m-2}+...+ \begin{cases}  {\bf Tr } \ \ \text{if} \ \ m \ \  \text{is} \ \ \text{even} \\  {\small \Yvcentermath1  \yng(1)} \ \ \ \text{if} \ \ m \ \ \text{is}  \ \ \text{odd}  \end{cases}
%\end{eqnarray}
%By using the complex variables introduced in \eqref{complex}, it is clear that $m=\bar Q$. Actually, only one operator of the irreducible representation $T_{\bar Q}$ carries a total charge equal to $\bar Q$, but the scaling dimension is the same for all the operators in the representation. Notice that all the operators of the form $T_{i_1...i_m}^{(m)}\left(\phi^2\right)^q$ transform in the same irreducible representation but clearly the one with the smallest scaling dimension has $q=0$.
%

The scaling dimension of $T_{\bar Q}$ at the fixed point has been computed to $\mathcal{O}(\epsilon^2)$ in \cite{Kehrein:1995ia}

 \begin{eqnarray}
\label{3loop}
     \Delta_{T_{\bar Q}} &&= \textcolor{red}{\bar Q+\left(-\frac{\bar Q}{2}+\frac{\bar Q(\bar Q-1)}{8+N}\right)}\epsilon-\left[\frac{184+N(14-3N)}{4(8+N)^3}\bar Q \right. \nonumber \\ && \left. +\textcolor{red}{\frac{(N-22)(N+6)}{2(8+N)^3}\bar Q^2}+\textcolor{red}{\frac{2}{(8+N)^2}\bar Q^3} \right]\epsilon^2 + \mathcal{O}\left(\epsilon^3 \right) 
 \end{eqnarray}

Terms highlighted in red will be used to test our results stemming from the semiclassical computation below. The scaling dimension to order $\mathcal{O}(\epsilon^3)$ has been calculated in \cite{Wallace:1974nu}, but we found out that it disagrees with the known literature for the special cases $\bar Q=2$ \cite{Braun:2013tva, Kompaniets:2019zes} and $N=2$ \cite{Badel:2019oxl}. In the next section, we will, therefore, combine our semiclassical results with the existing literature, in order to obtain the full $\mathcal{O}(\epsilon^4)$ anomalous dimension.

\section{Semiclassical approach to the $O(N)$ model} \label{computation}
We have now all the instruments to proceed with the computation of $\Delta_{T_{\bar Q}}$ semiclassically. 

\subsection{Classical contribution}

Here we focus on the leading term, $\Delta_{-1}$, which is given by the effective action \eqref{O2n_Lagrangian} evaluated on the classical trajectory \eqref{SolEOM} at the fixed point
\begin{equation}
  \frac{1}{g^*}\frac{\Delta_{-1} (g^* \bar Q )}{R}= \frac{\bar Q}{4}\left(3 \mu + \frac{m^2}{\mu}\right) \ .
\end{equation}
By inserting the second equation in \eqref{total} into the first one and setting $d=4$ we obtain
\begin{equation}
R^3 \mu^3 - R \mu = \frac{4}{3} \bar Q g^*
\end{equation}
with solution
\begin{equation}
  R \mu = \frac{3^\frac{1}{3}+\left(6 g^* \bar Q + \sqrt{-3+36 (g^* \bar Q)^2}\right)^\frac{2}{3}}{3^\frac{2}{3}\left(6 g^* \bar Q + \sqrt{-3+36 (g^* \bar Q)^2}\right)^\frac{1}{3}} \ .
\end{equation}
Thus the leading contribution is
\begin{align}
 \label{classic}
  \frac{4\Delta_{-1}}{g^* \bar Q} =  \frac{3^\frac{2}{3}\left(x+\sqrt{-3+x^2}\right)^{\frac{1}{3}}}{3^\frac{1}{3}+\left(x+\sqrt{-3+x^2}\right)^{\frac{2}{3}}}  + \frac{3^\frac{1}{3}\left(3^\frac{1}{3}+\left(x+\sqrt{-3+x^2}\right)^{\frac{2}{3}}\right)}{\left(x+\sqrt{-3+x^2}\right)^{\frac{1}{3}}}  \nonumber \\
   \end{align}
where $x \equiv 6 g^* \bar Q$. The expansion for small $g^* \bar Q$ reads
\begin{equation} \label{cl}
   \frac{\Delta_{-1}}{g^*} =   \bar Q \left[ 1 + \frac{1}{3}g^* \bar Q - \frac{2}{9}(g^* \bar Q)^2 + \frac{8}{27}(g^* \bar Q)^3 + \mathcal{O}\left((g^* \bar Q)^4\right)\right] \ .
\end{equation}
The leading term $\Delta_{-1}$ matches exactly the $U(1)$ result \cite{Badel:2019oxl}. This is a direct consequence of having a single chemical potential and it is consistent with the fact that the leading power of the charge at a given loop order in the perturbative expression for $\Delta_{T_{\bar Q}}$ does not depend on $N$. This can be easily seen by rewriting Eq.~\eqref{3loop} as a coupling expansion by mean of Eq~\eqref{WFFP}. Our result suggests that this behavior continues at higher loop orders.

\subsection{Quantum corrections}
The time is ripe to determine the leading quantum corrections $\Delta_0$ to be added to the classical result \eqref{classic}.
To this end, we expand around the saddle point configuration \eqref{SolEOM} considering the ground state in \eqref{rot}. We  parametrize the fluctuations as
\begin{equation}
\label{Parametrization}
  \begin{cases}
\chi_i =- i \mu t \ +\frac{1}{v}p_i(x) \,, & i=1,\dots,\frac{N}{2}-1 \,,\\
\chi_{N/2} =- i \mu t \ +\frac{1}{v}\pi(x) \,,\\
\sigma_i =  s_i(x)\,, & i=1,\dots,\frac{N}{2}-1 \,,\\
\sigma_{N/2} = v + r(x)\
\end{cases}
 \,,
\end{equation}
Expanding the Lagrangian \eqref{O2n_Lagrangian} to the quadratic order in the fluctuations, we arrive at 
\begin{eqnarray}
    \mathcal{L}_2 &&  =\frac{1}{2}(\partial \pi)^2+\frac{1}{2}(\partial r)^2 + (\mu^2-m^2)r^2-2 \ i \ \mu \ r \ \dot \pi \nonumber \\ && + \frac{1}{2} \partial s_i \partial s_i + \frac{1}{2} \partial p_i \partial p_i -2 \ i \ \mu \ s_i \ \dot p_i \ .
\end{eqnarray}
The spectrum for the non-abelian case contains states that are also seen in the abelian case, corresponding to one relativistic (Type I) Goldstone boson (the conformal mode) $\chi_{N/2}$ and one massive state $\sigma_{N/2}$ with mass $\sqrt{6 \mu^2-2 m^2}$. Their dispersion relations read 
\begin{equation}
    \omega_{\pm}(l) = \sqrt{J^2_\ell+3\mu^2-m^2 \pm \sqrt{4 J^2_\ell\mu^2+(3\mu^2-m^2)^2}} \ , 
\end{equation}
with the negative sign applying  to the Goldstone boson.
Additionally, the non-abelian case also features $n -1 = \frac{N}{2}-1$ non-relativistic (type II) Goldstone bosons $\chi_i$ and  $n - 1$ massive states $\sigma_i$ with mass $2\mu$
\begin{equation}
\omega_{\pm\pm}(l) = \sqrt{J^2_\ell + \mu^2} \pm \mu
\end{equation}
with $J^2_\ell = \ell (\ell+d-2)/R^2$ the eigenvalues of the Laplacian on the sphere.

The counting of Goldstone modes can be understood by recalling  that the symmetry breaking pattern is $U\left(\tfrac{N}{2}\right) \to U\left(\tfrac{N}{2}-1\right)$.  Naively one would have  expected $\dim \left(U\left(\tfrac{N}{2}\right)/U\left(\tfrac{N}{2}-1\right)\right) = N - 1$ relativistic Goldstone modes. However, the explicit Lorentz symmetry breaking due to the fixed charge modifies some of the Type I Goldstone bosons into fewer Type II  Goldstones. Each Type II counts as two Type I in the counting of the d.o.f with respect to the number of broken generators \cite{Nielsen:1975hm}. Thus we have
\begin{equation}
  1 + 2\times \left( \frac{N}{2}-1 \right) = N - 1 = \dim \left(U\left(\tfrac{N}{2}\right)/U\left( \frac{N}{2}-1 \right)\right) \,.
\end{equation}
$\Delta_0$ is determined by the fluctuation functional determinant and it is given by
\begin{equation}
\label{eq:one-loop-det1}
\Delta_0 = \frac{R}{2}\sum_{\ell=0}^\infty n_{\ell}\left[\omega_+(\ell)+\omega_-(\ell)+(\tfrac{N}{2}-1)(\omega_{++}(\ell)+\omega_{--}(\ell))\right]\, 
\end{equation}
where $n_\ell=(1+\ell)^2$ is the Laplacian multiplicity on the $3$-sphere.

The difference with respect to the abelian case is that now we have to include the contributions of all the $2 \times \left(\tfrac{N}{2}-1\right)$ new modes.
As a result, the rank of the $O(N)$ group $n$ enters in the computation and leads to a non-trivial dependence of the leading quantum corrections on the number of scalars $N$.

Following the procedure of \cite{Badel:2019oxl}, we obtain
\begin{eqnarray}
\Delta_0(g^* \bar Q)= && -\frac{15 \mu^4  R^4+6 \mu^2  R^2-5}{16}
+\frac{1}{2} \sum_{\ell=1}^\infty\sigma(\ell)
+\frac{\sqrt{3\mu^2R^2-1}}{\sqrt{2}} \nonumber \\ && -\frac{1}{16}\left(\frac{N}{2}-1\right)\left[7 + R \mu \left(-16+6R\mu+3R^3 \mu^3\right)\right] \ .
\label{Delta0}
\end{eqnarray}
The last term and sum $\sigma(\ell)$  distinguish the non-abelian case from the abelian one with 
\begin{widetext}
\begin{align}
\sigma(\ell) &=  R (1+\ell)^2\left[\omega_+(\ell)+\omega_-(\ell)+(\tfrac{N}{2}-1)(\omega_{++}(\ell)+\omega_{--}(\ell))\right]+\frac{1}{8}(N+8)\left( R^2 \mu ^2 -1\right)^2\frac{1}{\ell} \nonumber \\
& +\frac{1}{2}\left(2-N-(N+2) R^2 \mu^2 \right)+\frac{1}{2}\left(2-5N-(2+N) R^2 \mu^2 \right) \ \ell -3 N \ \ell^2- N \ \ell^3 \ .
\end{align}
where all the quantities are evaluated in $d=4$ dimensions.  The above constitute our main achievement.

As a non-trivial test of our result,  we now compare it with the  perturbative results (the red contributions in \eqref{3loop}) to the maximum known order in perturbation theory. To do this we  expand the result for small $g \bar Q$, where the sum above can be computed analytically. 

\begin{eqnarray} \label{1loop}
    \Delta_0(g^* \bar Q)=  -\left(\frac{5}{3}+\frac{N}{6} \right) g^* \bar Q +\left(\frac{1}{3}-\frac{N}{18} \right) \ (g^* \bar Q)^2  +\frac{1}{27}[N-36+28 \  \zeta(3)+2N \ \zeta(3)] \ (g^* \bar Q)^3 +\mathcal{O}\left( (g^* \bar Q)^4 \right) \ . 
    \end{eqnarray}

The sum of the classical contribution \eqref{cl} and the leading quantum correction \eqref{1loop} is

\begin{eqnarray}
\frac{\Delta_{-1}(g^* \bar Q)}{g^*}+\Delta_0(g^* \bar Q)=  \bar Q -\frac{g^* \bar Q}{6}(10+N-2{\bar Q})   + \frac{{g^*}^2\bar Q^2}{18}(6-N-4\bar Q )  +\frac{{g^*}^3\bar Q^3}{27} [N-36+8\bar Q+2(14+N) \zeta(3)]+\mathcal{O}( g^{*4} \bar Q^5) \ . \nonumber \\
\end{eqnarray}

To better visualise our results we now rewrite the above at the WF fixed point \eqref{WFFP} so that directly compare with the red terms of $\eqref{3loop}$ as follows
%\begin{widetext}
\begin{eqnarray}
\label{3loopred}
   {\color{red} \Delta_{T_{\bar Q}}}  &&=\bar Q+\left(-\frac{\bar Q}{2}+\frac{\bar Q(\bar Q-1)}{8+N}\right)\epsilon-\left[ \frac{(N-22)(N+6)}{2(8+N)^3}\bar Q^2+\frac{2}{(8+N)^2}\bar Q^3 \right]\epsilon^2  \ .
 \end{eqnarray}
%\end{widetext}
At order $\cO(\epsilon^3)$ and $\cO(\epsilon^4)$, our semiclassical computation captures the leading and next to leading terms in the charge. However, we can determine also the remaining terms by asking that we reproduce the known results for the cases $\bar Q = 1$ \cite{Kleinert:1991rg}, $\bar Q = 2$ \cite{Braun:2013tva, Kompaniets:2019zes}, and $\bar Q = 4$ \cite{Calabrese:2002bm, Johan}. In this way, we obtain the full $4$-loop scaling dimension of $T_{\bar Q}$, which reads
%\begin{widetext}
 \begin{eqnarray}
\label{3loopcomplete}
    && \Delta_{T_{\bar Q}}=\bar Q+\left(-\frac{\bar Q}{2}+\frac{\bar Q(\bar Q-1)}{8+N}\right)\epsilon-\left[\frac{184+N(14-3N)}{4(8+N)^3}\bar Q+\frac{(N-22)(N+6)}{2(8+N)^3}\bar Q^2+\frac{2}{(8+N)^2}\bar Q^3 \right]\epsilon^2 + \left[\frac{8}{(8+N)^3}\bar Q^4 \right. \nonumber\\
    && +\frac{-456-64N+N^2+2(8+N)(14+N)\zeta(3)}{(8+N)^4}\bar Q^3 +\frac{-N^4-57 N^3+258 N^2-24 (N+6) (N+8) (N+26) \zeta (3)+8176 N+31008}{4 (N+8)^5}\bar Q^2 \nonumber \\
    &&\left. +\frac{-69504 + 3 N [-5216+ N (184+ N(86+N))]+64 (8+N)(178+ N (37 + N))\zeta(3)}{16
   (N+8)^5}\bar Q\right]\epsilon^3 + \left[-\frac{42}{(8+N)^4} \bar Q^5 \right. \nonumber \\
    &&\left. +\frac{-4 N^2-5 (N+8) (N+30) \zeta(5) -2 (N+8) (6 N+65) \zeta(3) +476
   N+3344}{(8+N)^5} \bar Q^4 + \frac{1}{60 (8+ N)^6} \left(\pi^4 N^4+60 N^4+38 \pi^4 N^3+4020 N^3 \right. \right. \nonumber \\
    &&\left. \left.+528 \pi^4 N^2-88800 N^2-4200 (N-2) (N+8)^2 \zeta (5)-60 (N+8) (N (N (3
   N-44)-1720)-7464) \zeta(3) +3200 \pi^4 N-1577280 N \right. \right. \nonumber \\
    &&\left. \left. +7168 \pi^4-5662560 \right) \bar Q^3 - \frac{1}{80 (N+8)^7} \left( 10 N^6+4 \pi ^4 N^5+915 N^5+224 \pi^4 N^4+34120 N^4+4464 \pi^4
   N^3+86600 N^3+41600 \pi ^4 N^2 \right. \right. \nonumber \\
    &&\left. \left. -3928440 N^2-400 (N+8)^2 (N (65
   N+958)+2496) \zeta (5)-20 (N+8) (N (N (N (N
   (N+52)+904)-12224)-181184)-514112) \zeta (3) \right. \right. \nonumber \\
    &&\left. \left. +185344 \pi^4  N-35161600 N+319488 \pi^4 - 87127680\right) \bar Q^2 +\frac{1}{960 (8+ N)^7} \left(45 N^6+32 \pi ^4 N^5+5820 N^5+1952 \pi^4 N^4+322440 N^4  \right. \right. \nonumber \\
    &&\left. \left. +40256
   \pi^4 N^3+1972440 N^3+380416 \pi ^4 N^2-16196640 N^2-9600
   (N+8)^2 (N (25 N+418)+1240) \zeta (5)  \right. \right. \nonumber \\
    &&\left. \left. -240 (N+8) (N (N (N (N
   (N+40)+1056)-3496)-100480)-300096) \zeta (3)+ 1699840 \pi ^4 N -191091840 N+2916352 \pi^4  \right. \right. \nonumber \\
    &&\left. \left. -494461440  \big) \bar Q \Bigg]\epsilon^4  +  \mathcal{O}\left(\epsilon^5 \right)  \right. \right. \ .
 \end{eqnarray}
%\end{widetext}
As anticipated, the above result corrects the one in \cite{Wallace:1974nu} demonstrating the power of the approach. Moreover, we can now predict the classical and quantum correction for the higher perturbative loops of the anomalous dimension. To help future checks we give the explicit result up to order ${g^*}^6$. 
\begin{eqnarray}
\text{5-loops:} && \left(\tfrac{256}{243} \bar Q+ \tfrac{1}{243}[3(-800+7 N) + 28  \zeta(3) (28+ 3 N) +40  \zeta(5)(22 + N) +14 \zeta(7)(62+ N)]\right)(g^* \bar Q)^5 \\
\text{6-loops:} && \left(-\tfrac{572}{243}\bar Q +\tfrac{2}{279}[10191-64 N - 2 \zeta(3)(1327+160 N) - 2 \zeta(5)(1441 + 80 N)-70 \zeta(7) (46 +N)- 21 \zeta(9)(126 +N)]\right)(g^* \bar Q)^6 \nonumber \\
\end{eqnarray}
\end{widetext}

Since the result is valid for any $N$ one can now directly apply it to the Higgs sector of the Standard Model for which $N=4$ and up to Yukawa and gauge interactions. 

In this letter we focussed on the limit $g^\ast \bar{Q}$ fixed and small. This allowed us to determine the all order perturbative contributions to the quantum scaling dimension related to the fixed charge. Another interesting and complementary limit,  already much explored in the literature \cite{Alvarez-Gaume:2016vff}, is the one in which $g^\ast \bar{Q}$ is large  that can be straightforwardly obtained from \eqref{classic} and \eqref{Delta0}.

\smallskip
Concluding, we used the semiclassical approach to determine, for the first time, the scaling dimensions for the critical $O(N)$ model in $4-\epsilon$ dimensions.  We determined the scaling dimension to the leading order and next to leading order terms in the  charge expansion but to all-orders in the coupling.  This work constitutes the stepping stone towards generalising the approach to gauge-Yukawa theories.  

After this work was finished, a related study in $d=6-\epsilon$ appeared in \cite{Arias-Tamargo:2020fow}. 

The work of O.A. and J.B. is partially supported by the Croatian Science Foundation project number 4418 as well as European Union through the European Regional Development Fund - the Competitiveness and Cohesion Operational Programme (KK.01.1.1.06). F.S and Z.W acknowledge the partial support by Danish National Research Foundation grant DNRF:90. We would like to thank Johan Henriksson for his valuable comments.

  \end{document}